\documentclass[journal=nalefd,manuscript=letter]{achemso}
\pdfoutput=1 
\usepackage{graphicx}

\author{Paloma A. Huidobro}
\author{Maxim L. Nesterov}
\altaffiliation{A. Ya. Usikov Institute for Radiophysics and
Electronics, NAS of Ukraine, 12 Academician Proskura Street, 61085
Kharkov, Ukraine} \affiliation[Departamento de F\'isica Te\'orica
de la Materia Condensada, Universidad  Aut\'onoma de
Madrid]{Departamento de F\'isica Te\'orica de la Materia
Condensada, Universidad Aut\'onoma de Madrid, E-28049 Madrid,
Spain}
\author{Luis Mart\'in-Moreno}
\affiliation[Instituto de Ciencia de
Materiales de Arag\'on and Departamento de F\'isica de la Materia Condensada,
CSIC--Universidad de Zaragoza]{Instituto de Ciencia de
Materiales de Arag\'on and Departamento de F\'isica de la Materia Condensada,
CSIC--Universidad de Zaragoza, E-50009 Zaragoza, Spain}
\author{Francisco J. Garc\'ia-Vidal}
\email{fj.garcia@uam.es} \affiliation[Departamento de F\'isica
Te\'orica de la Materia Condensada,  Universidad Aut\'onoma de
Madrid]{Departamento de F\'isica Te\'orica de la Materia
Condensada, Universidad Aut\'onoma de Madrid, E-28049 Madrid,
Spain}
\title[]{Transformation Optics for Plasmonics}

\begin{document}
\begin{abstract}
A new strategy to control the flow of surface plasmon polaritons
at metallic surfaces is presented. It is based on the application
of the concept of Transformation Optics to devise the optical
parameters of the dielectric medium placed on top of the metal
surface. We describe the general methodology for the design of
Transformation-Optical devices for surface plasmons and analyze,
for proof-of-principle purposes, three representative examples
with different functionalities: a beam shifter, a cylindrical
cloak and a ground-plane cloak.
\end{abstract}

\newpage
Transformation Optics (TO) has been proposed \cite{PendrySci06,
Leonhardt06} as a general technique to design complex
electromagnetic (EM) media with unusual properties. In order to
construct novel optical devices, one has to imagine a space with
some distortion in a certain region and make a coordinate
transformation from a flat empty space. Due to the form-invariance
of Maxwell equations under general coordinate transformations
\cite{Pendry96}, the properties of the distorted geometry can then
be interpreted as that of a medium in the original flat empty
space  \cite{LeonhardtPhilbin06}. Thus, TO
provides us with expressions for the dielectric permittivity
tensor,  $\hat{\epsilon}$, and the magnetic permeability tensor,
$\hat{\mu}$, that need to be implemented in order to obtain a
medium with a designed functionality. The first practical
realization of this idea was the construction of a two-dimensional
(2D) cylindrical invisibility cloak for EM waves in the microwave
regime \cite{Schuring06}. This experiment took advantage of the
recently developed field of Metamaterials\cite{Ziolkowski,Ramakrishna}. These are artificially
structured materials made up of subwavelength constituents and
designed to implement a prescribed response to EM fields. The
cylindrical cloak experiment was followed by the construction of
broad-band cloaks in the microwave \cite{Liu09} and optical \cite{Valentine09, Gabrielli} regimes, all of them based on a different strategy, the so-called ground-plane cloak \cite{JensenLi08}. Besides, a wide variety of
applications other than cloaking has been recently presented,
including beam shifters and beam dividers \cite{Rahm08}, field
concentrators \cite{RahmSchuring08}, collimators \cite{Kildishev}, waveguide bends and corners
\cite{Smith08}, lenses with subwavelength resolution
\cite{Schuring07} or deep-subwavelength waveguides \cite{Han08}.

In Plasmonics, one of the main goals is to control the flow of
light at a metal surface by means of the surface plasmon
polaritons (SPPs) that decorate a metal-dielectric interface \cite{Barnes03,Maier05}. In
the quest for creating photonic circuits based on SPPs, different
waveguiding schemes have been tested during the last years (see,
for example, the recent review articles
\cite{Bozhevolnyi08,Gramotnev10}). At a different level, 2D
optical elements for SPPs like mirrors and beam splitters based on
Bragg-scattering of SPPs with periodic arrays of scatterers have
been also devised \cite{Ditlbacher02}.

In this paper we present a different strategy to tackle this
problem by showing how the TO framework can be applied to mold the
propagation of SPPs at a metal surface. As TO is valid for any
source of illumination due to the universal character of Maxwell
equations, it is expected that TO recipes will be also operative
for SPPs. These EM modes are surface waves travelling along the
interface between two media, a dielectric and a metal. Therefore,
in order to design an optical device for SPPs, the expressions for
the EM material parameters provided by the TO formalism should, in
principle, be implemented both in the dielectric and the metal
sides. This fact is a severe technical challenge, as TO generally
requires highly anisotropic and inhomogeneous $\hat{\epsilon}$ and
$\hat{\mu}$. Moreover, the decay length of the SPP within the
metal, i.e. the skin depth, is subwavelength, which means that a
manipulation of $\hat{\epsilon}$ and $\hat{\mu}$ at a nanometer
scale within the metal would be needed. However, we will show in
the following that a simplified version of the TO recipe in which
only the dielectric side is manipulated leads to quasi-perfect
functionalities. Two factors determine whether this simplification
is operative or not: the wavelength of the SPP and the geometry of
the device under consideration. In order to illustrate this point
we discuss here three very different devices: a beam shifter, a
cylindrical cloak and a ground-plane cloak for SPPs.

As a first illustration of our methodology, we consider the 3D
implementation for SPPs of a parallel beam shifter recently
proposed by Rahm and co-workers for a 2D geometry \cite{Rahm08}.
Let us discuss first this 2D case that was the first report of a
finite embedded transformation, which is capable of transferring
the modification of the EM fields to the wave that exits the TO
medium. In the design of the 2D beam shifter, a rectangular region
of sides $d$ and $l$ is considered and a transformation is
performed from a 2D Cartesian grid into a grid that is tilted at
an angle $\phi$ [see inset of Figure 1(a)]. The amount of shift
that the incident beam will experience is given by $a=\tan\phi$,
which we will set to 1 ($\phi=45^0$) throughout this work for
simplicity. The symmetry of the 2D problem for a EM plane wave
implies that the two polarizations can be treated separately. For
a wave polarized with the electric field perpendicular to the
plane of the transformation, a TM plane wave, the relevant EM
parameters obtained for the transformation medium according to TO
are $\epsilon_{zz}=1$, $\mu_{xx}=1$, $\mu_{xy}=\mu_{yx}=a$ and
$\mu_{yy}=1+a^2$.

To design a 3D beam shifter for a SPP propagating along the
interface between a metal and a dielectric, we perform the
embedded transformation in parallelepipeds of rectangular base ($d
\times l$). Two different transformation media are needed, one in
the dielectric, with height $h_d$, and one in the metal (height
$h_m$). Notice that in order to operate over the entire SPP field,
the height of the rectangular slab in each medium must be greater
than the corresponding decay length of the SPP in that medium
(metal or dielectric). The expressions for $\hat{\epsilon}$ and
$\hat{\mu}$ inside the dielectric slab in a cartesian basis are
independent of the $z$ coordinate and are:

\begin{equation}
\label{eq:BSparameters1}
\hat{\epsilon}_{TO}=\hat{\mu}_{TO}=
\left( \begin{array}{ccc}
    1&a&0\\a&1+a^2&0\\0&0&1
\end{array}\right)
\end{equation}
where we have assumed that the dielectric half-space is vacuum
without any loss of generality. On the other hand, the two tensors
in the metallic slab should be:

\begin{equation}
\label{eq:BSparameters2}
\hat{\epsilon}=\hat{\epsilon}_{TO}\cdot\epsilon_m \textrm{, }\ \hat{\mu}=\hat{\mu}_{TO}
\end{equation}
where $\epsilon_m$ is the electric permittivity of the metal.

\begin{figure}[h]
\includegraphics[width=8.5cm]{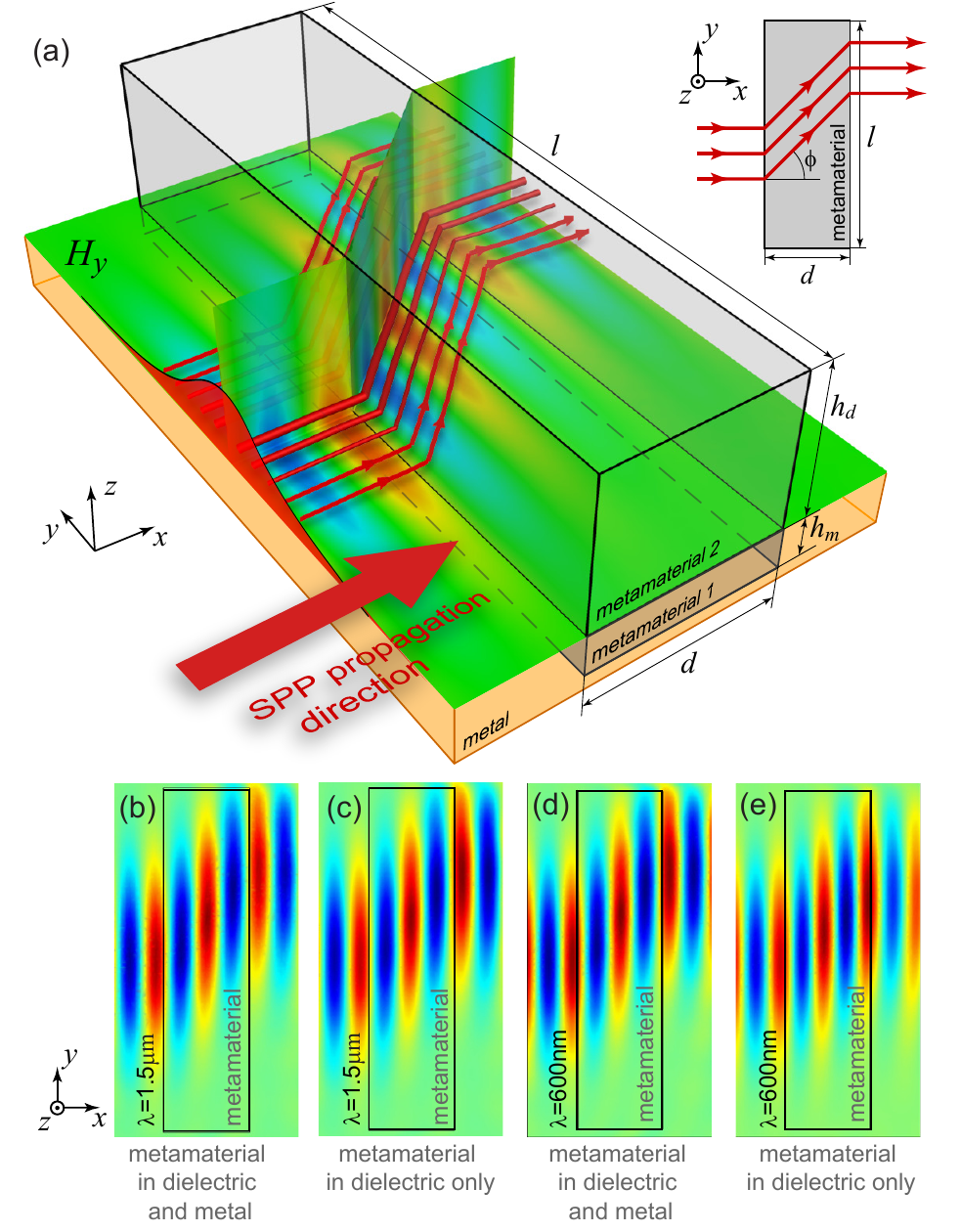}
\caption{Normalized magnetic field pattern (color map) of a SPP
travelling along a gold-vacuum interface and entering into a 3D
beam shifter. Inset of the panel (a) shows the basic geometry of
the 2D case. Panel (a) depicts a 3D view with power flow
streamlines of a SPP operating at $\lambda=1.5$ $\mu$m. The
prescribed $\hat{\epsilon}$ and $\hat{\mu}$ were implemented in
rectangular parallelepipeds with in-plane dimensions $2.5 \mu$m
$\times 10\mu$m and thicknesses $h_d=6$ $\mu$m and
$h_m=50$ nm. Panel (b) renders a cross-section picture of panel
(a) evaluated at the metal surface, $z=0$. For comparison, panel
(c) shows a SPP beam shifter where the slab with the prescribed
$\hat{\epsilon}$ and $\hat{\mu}$ has been introduced only in the
dielectric side. Panels (d) and (e) render the operation of a beam
shifter for a SPP of $\lambda=600$ nm. In this case the in-plane
dimensions of the rectangular slabs are $1 \mu$m $\times 4\mu$m
and the thicknesses are $h_d=600$ nm and $h_m=50$ nm. In panel
(e), only the dielectric side is manipulated.}
    \label{fig:BeamShifter}
\end{figure}

To test the TO recipe for a SPP beam shifter, we have carried out
3D full-wave simulations using the COMSOL Multiphysics software
that is based on the finite element method (FEM).
\ref{fig:BeamShifter}(a) shows a 3D view of the behaviour of a SPP
at $\lambda=1.5$ $\mu$m going through a beam shifter. The incident
SPP, which displays a gaussian profile in the $y$-direction,
propagates along the $x$-direction through a gold-vacuum interface
and enters into the transformation medium, where it is shifted
in the $y$-direction. Details on the geometrical parameters can
be found in the caption of \ref{fig:BeamShifter}. In this
simulation, rectangular slabs with the prescribed $\hat{\epsilon}$
and $\hat{\mu}$ were introduced in both the dielectric and metal
sides. A cross-section showing the behaviour of $H_y$ at the
metal surface ($z=0$) is rendered in \ref{fig:BeamShifter}(b).
\ref{fig:BeamShifter}(c) and (b) allow us to compare the quality
of the shift whether we implement $\hat{\epsilon}$ and $\hat{\mu}$
in the dielectric and metal sides (b) or only in the dielectric
side (c). It is worth noticing that there is almost no difference
between the field distributions in both cases. On the other hand,
\ref{fig:BeamShifter}(d) and (e) show the corresponding
cross-sections at $z=0$ of the beam shifter for a SPP but now
operating at $\lambda=600$ nm for the case of placing the
transformation media in the dielectric and metal sides (d) and
when only the dielectric side is manipulated (e). Although the
effectiveness of the shifter at this wavelength is affected by the
absence of a transformation medium in the metal side, its
functionality (a shift in the SPP beam) is still very good. The
reason for this weak wavelength dependence is clear. The vacuum
decay length of a SPP on a gold surface at $1.5$ $\mu$m is $2.4$ $\mu$m, much larger
than the skin depth ($23$ nm) so the EM fields within the metal carry 0.01$\%$ of the total SPP energy, 
having practically no relevance compared to the EM fields in the
dielectric. However, at $\lambda=600$ nm, the ratio between vacuum
decay length of the SPP and skin depth is reduced by a factor of
$10$ and the EM fields inside the metal are then relatively more
important for the propagation of the SPP. For this wavelength, 1$\%$ of the SPP energy resides in the metal. 

As a second example, we present a 3D cylindrical cloak for SPPs
based on TO. Here we aim for suppressing the scattering of a SPP from an object placed on a surface and not to use plasmonic structures to achieve cloaking \cite{Engheta05}. Note that other strategies for SPP cloaking that do
not rely on the TO framework have been recently proposed
\cite{Maradudin}. As stated above, the 2D invisibility cloak was
the first application of TO, both in theoretical proposals
\cite{PendrySci06, Leonhardt06} and experiments \cite{Schuring06}.
The design of a 3D cylindrical cloak for SPPs is analogous to that
of the beam shifter. The field distribution of the SPP extends
itself in the dielectric and in the metal, so, in principle, we
should again cloak the SPP-field in both regions of space. Because
of the presence of the metal-dielectric interface, we apply the 2D
cylindrical transformation as described by Pendry and co-workers
\cite{PendrySci06} in planes that are parallel to the interface,
building-up two concentric cylinders of radii $a$ and $b$ [see
Figure 2(a)]. In each plane, we perform a transformation from a flat 2D
euclidean space, where light propagates in straight lines, to a
space with a hole of radius $a$ surrounded by a compressed region
of radius $b$. The transformation medium is the shell comprised
between $a$ and $b$ and is expected to guide the
incident SPP around the hole. The SPP-wave will emerge at the
other side as if it had travelled through an empty space. The
$\hat{\epsilon}$ and $\hat{\mu}$ tensors obtained for the transformation medium in the dielectric
half-space read as follows in a cylindrical basis:

\begin{equation}
\label{eq:CylCparameters}
\hat{\epsilon}_{TO}=\hat{\mu}_{TO}= \left( \begin{array}{ccc}
    \frac{r-a}{r}&0&0\\0&\frac{r}{r-a}&0\\0&0&\left(\frac{b}{b-a}\right)^2\frac{r-a}{r}
\end{array}\right)
\end{equation}
Cloaking the SPP field inside the metal would require a
cylindrical transformation medium whose magnetic permeability is
the same as in the dielectric side, $\hat{\mu}_{TO}$,  while its electric permittivity
is simply $\hat{\epsilon}_{TO} \cdot \epsilon_m$.

\begin{figure}[h]
\includegraphics[width=8.5cm]{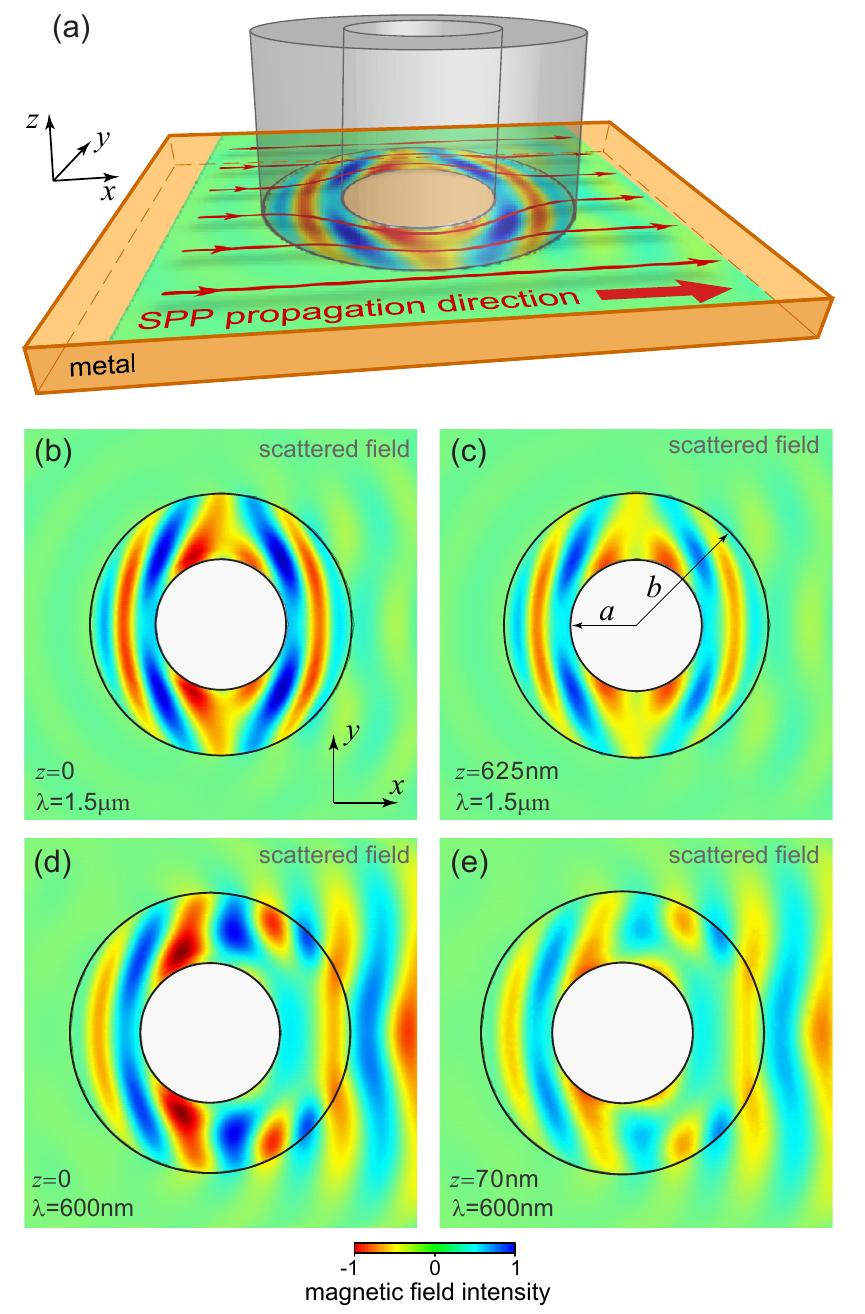}
\caption{Scattered magnetic field for a 3D cylindrical cloak for
SPPs. (a) A schematic picture showing the geometry of the cloak.
Cloaking parameters are implemented only in the dielectric side,
in the region between the inner radius $a$ and the outer radius
$b=2a$. The operating wavelength is $1.5 \mu$m in panels (b) and
(c) and the cylindrical cloak in this case is $6\mu m$ high with
$a=1.5\mu$m. (b) Scattered $H_y$ field evaluated at the metal
surface, $z=0$. (c) The corresponding field as obtained at $z=625$
nm, a quarter of the vacuum decay length of the SPP into the
dielectric. In panels (d) and (e) the wavelength is $600$ nm and
the cloak is $600$ nm high with $a=500$ nm. (d) Scattered $H_y$
field obtained at $z=0$ plane. (d) The same quantity but now
evaluated at $z=70$ nm, a quarter of the vacuum decay length of
the SPP for $\lambda=600$ nm.}
    \label{fig:CylCloak}
\end{figure}

3D simulations of the cylindrical cloak for SPPs operating at
$\lambda=1.5$ $\mu$m [panels (b) and (c)] and $\lambda=600$ nm
[panels (d) and (e)] are shown in \ref{fig:CylCloak}. In these
cases, the prescribed $\hat{\epsilon}$ and $\hat{\mu}$ were
implemented only in the dielectric, in a cylindrical shell
surrounding a perfectly conducting cylinder, leaving the metal as
a continuous sheet. We present for each wavelength the
$y$-component of the scattered magnetic field evaluated at two
planes parallel to the metal-dielectric interface, one at $z=0$
[panels (b) and (d)] and another at a height at which the SPP has
decayed to a quarter of its intensity at the interface [panels (c)
and (e)]. At $\lambda=1.5$ $\mu$m, the scattered magnetic field is
nearly zero out of the cloaking region and we can conclude that the
performance of the cloaking device for SPPs is excellent. However,
for a SPP at a wavelength of $600$ nm, only partial cloaking for a
SPP is achieved. The reason for this is the geometry of the
cylindrical cloak which, as opposed to the beam shifter, has a
discontinuity in the metal-dielectric interface. In this case, the
behaviour of the cloak is very sensitive to the field propagating
inside the metal because it strongly scatters at this
discontinuity. Simulations in which the metal is also manipulated
according to the TO recipe were carried out (not shown here). In
this case, the 3D cylindrical cloak for SPPs displays a perfect
performance at the two wavelengths considered. Our results explain why a previous experiment on SPP cloaking at $\lambda=500$ nm in which the metamaterial was only placed in the dielectric side was not completely successful \cite{Smolyaninov}.

Finally, we consider a different strategy for cloaking
\cite{JensenLi08}, based on the use of TO to design a cloak that
mimics a ground-plane (flat metal surface). We will show that in
this case the simplification of using the transformation media
only in the dielectric side works perfectly. The main difference
between a ground-plane cloak and the rest of the TO-based devices
that we have considered up to here is the geometry. A ground-plane
cloak inherently needs to be placed on a metal surface, because it
mimics a highly reflective surface instead of an empty space.
Taking advantage of the presence of this metal-dielectric
interface, we apply directly the TO recipe previously derived for
a 2D ground-plane cloak illuminated by a plane wave
\cite{JensenLi08} to the case of SPP-illumination.

\begin{figure}[h]
\includegraphics[width=8.5cm]{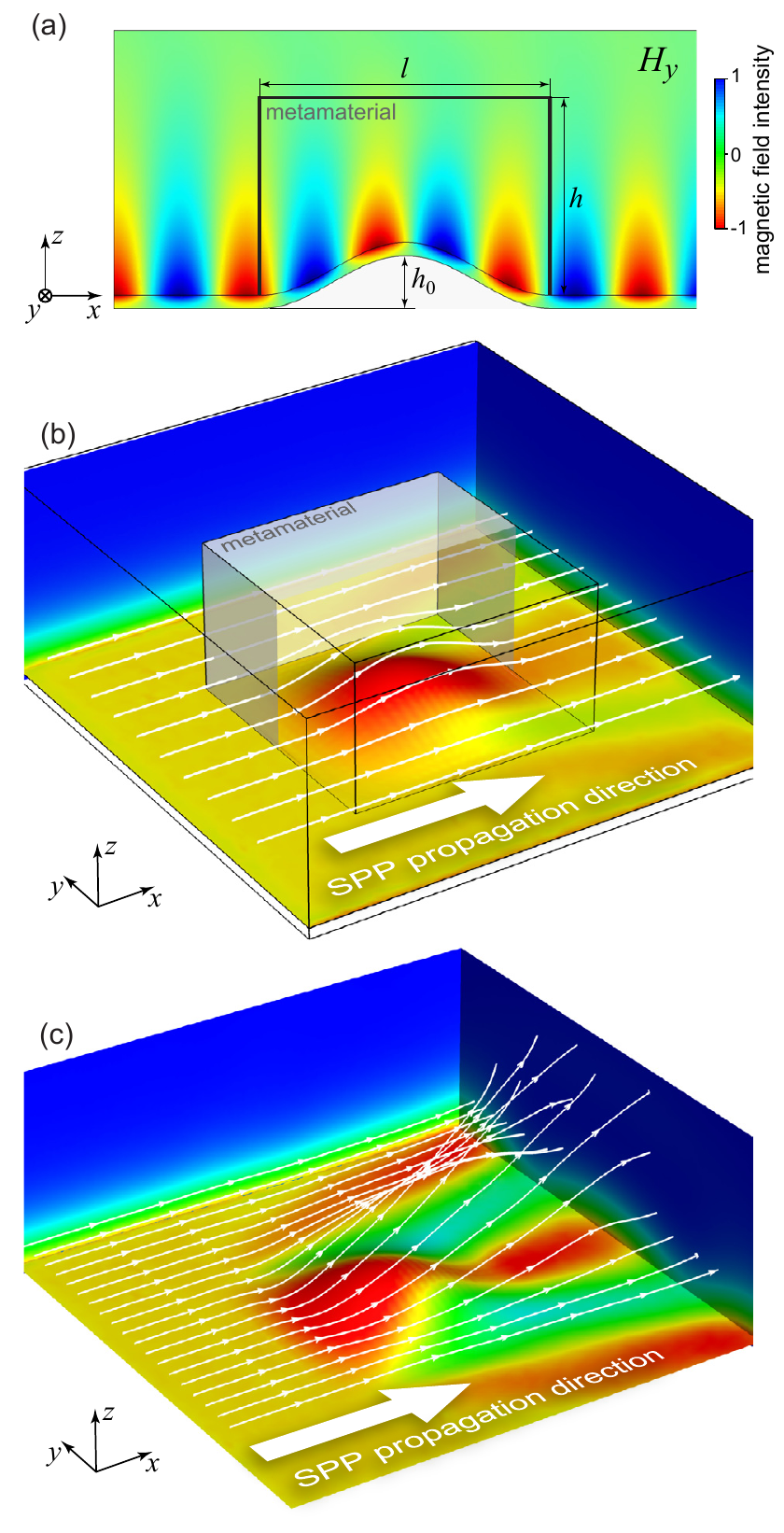}
\caption{(a) Normalized magnetic field pattern of a 2D
ground-plane cloak for a SPP travelling in the $x$-direction at
$600$ nm. The cloaked object is a $\cos^2$-shaped gold bump,
$1250$ nm long and with a height of $h_0=200$ nm. The height of
the cloak is $h=750$ nm. (b) Power flow distribution (color map)
and streamlines of a 3D ground-plane cloak placed over a 3D
$\cos^2$-shaped gold bump of dimensions $1250$ nm $\times$ $1250$
nm $\times$ $200$ nm. (c) Power flow for the 3D gold scatterer
without the cloak.}
    \label{fig:CarpetCloak}
\end{figure}

The design of a 2D ground-plane cloak \cite{JensenLi08} starts by
considering a rectangular region of a flat space. The size of the
rectangle is $l \times h$ and it sits on a metal surface. Suppose
that we want to cloak a bump of maximum height $h_0$ and with a
shape that follows the equation $z(x)=h_0
\cos^2\left(\frac{\pi}{l}x\right)$. Then we make a transformation
from the rectangle to a region of the same shape except for its
bottom boundary which is lifted following the surface of the bump.
The simplest transformation consists of a transfinite mapping,
which is not conformal \cite{Knupp94}. This means that it does not
preserve angles and hence the anisotropy is not minimized
\cite{JensenLi08}. However, we will use it as a proof-of-concept
of the performance of a ground-plane cloak for SPPs. If we place
the cloak in the $x-z$ plane (see \ref{fig:CarpetCloak}), the EM
fields along the $y$-axis being translationally invariant, the
transformation in a cartesian basis reads:

\begin{equation}
\label{eq:2DCarpetCloakTransformation}
 x=x' \textrm{,} \ y=y' \textrm{,} \ z=z'+h_0\left(1-\frac{z'}{h}\right)\cos^2\left(\frac{\pi}{l}x'\right)
\end{equation}
The optical parameters, $\hat{\epsilon}$ and $\hat{\mu}$, for the transformation medium
are obtained using the standard TO techniques:

\begin{equation}
\label{eq:2DCarpetCloakParameters}
\hat{\epsilon}=\hat{\mu}= \frac{1}{\Delta}\left( \begin{array}{ccc}
1&0&f_{xz} \\ 0 & 1& 0 \\ f_{xz}& 0 & f_{zz}
    \end{array} \right)
\end{equation}
where $\Delta=\frac{\partial z}{\partial z'}$,
$f_{xz}=\frac{\partial z}{\partial x'}$ and
$f_{zz}=f_{xz}^{\phantom{yz}2}+\Delta^2$. The 2D
ground-plane cloak for SPPs consists of a dielectric with the
prescribed $\hat{\epsilon}$ and $\hat{\mu}$ placed on a metal
surface with a $\cos^2$-shaped bump. A SPP propagating along the
metal-dielectric interface and entering into the ground-plane
cloak should be smoothly guided around the bump.
\ref{fig:CarpetCloak}(a) shows the magnetic field distribution of
a SPP at $\lambda=600$ nm travelling along a gold-vacuum interface
and going through the cloak. Although only the dielectric medium
is transformed in this 2D simulation, the SPP nicely follows the
curvature of the bump at the two sides of the interface.

A ground-plane cloak for SPPs can be also devised for hiding a 3D
bump. In this case, the bump sits on the $x-y$ plane and its
surface is given by $z(x,y)=h_0
\cos^2\left(\frac{\pi}{l}x\right)\cos^2\left(\frac{\pi}{l}y\right)$.
As for the 2D cloak, we make a transfinite transformation from a
rectangular prism to a prism whose bottom boundary is lifted by
the bump. In this case, the mathematical formulae for the
transformation are:

\begin{equation}
\label{eq:3DCarpetCloakTransformation}
x=x'\ \textrm{,} \ y=y'\ \textrm{,} \ z=z'+h_0\left(1-\frac{z'}{h}
\right)\cos^2\left(\frac{2\pi}{l}x'\right)\cos^2\left(\frac{2\pi}{l}y'\right)
\end{equation}
According to the TO rules, we obtain the $\hat{\epsilon}$ and
$\hat{\mu}$ for the transformation medium:

\begin{equation}
\label{eq:3DCarpetCloakParameters}
    \hat{\epsilon}=\hat{\mu}=
   \frac{1}{\Delta}
        \left( \begin{array}{ccc}
       1 & 0 & f_{xz} \\
         0 & 1 & f_{yz} \\
        f_{xz}&  f_{yz} & f_{zz}
    \end{array} \right)
\end{equation}
where $\Delta=\frac{\partial z}{\partial z'}$,
$f_{xz}=\frac{\partial z}{\partial x'}$, $f_{yz}=\frac{\partial
z}{\partial y'}$ and $f_{zz}=
f_{xz}^{\phantom{yz}2}+f_{yz}^{\phantom{yz}2}+\Delta^2$.
We present in \ref{fig:CarpetCloak}(b) a simulation that confirms
the effectiveness of the 3D cloak for a SPP operating at
$\lambda=600$ nm. The color map and the streamlines represent the
density and the direction of the power flow of the SPP,
respectively. The SPP is guided around the bump and continues
travelling along the gold-vacuum interface. For comparison,
\ref{fig:CarpetCloak}(c) shows a SPP scattering from the gold bump
when the cloak is not present. We see that the fraction of the SPP
that impinges directly into the bump is scattered upwards. This
radiation into free space is totally suppressed when the cloak is present, as shown in \ref{fig:CarpetCloak}(b).

Now we briefly discuss the practical realization of the
TO-based devices for SPPs presented in this work. Our results have
shown that the effectiveness of these devices is basically
governed by the presence of a transformation medium in the
dielectric side whereas modifying the metal side is not necessary
in most of the cases. Then it is enough to realize anisotropic
$\hat{\epsilon}$ and $\hat{\mu}$ within the dielectric, making use of a specific metamaterial. Different tools for overcoming the necessity of anisotropic parameters have already been developed within the field
of Metamaterials. For instance, in the reduced parameters approach, the
tensors are simplified so that only some components need to be
implemented. This allows the design of non-magnetic metamaterials operating in
the optical regime \cite{Shalaev}, a fact that has been applied to the construction of the 2D cylindrical cloak. On the other hand, isotropic $\hat{\epsilon}$ and $\hat{\mu}$ are obtained when a conformal transformation\cite{Leonhardt06} is performed, so it is enough to implement an isotropic refractive index. Numerical techniques have been recently developed to compute quasi-conformal coordinate transformations, which minimize anisotropy to a point where it is negligible\cite{JensenLi08}. In particular, the quasi-conformal mapping technique has proved useful for the
practical realization of the 2D ground-plane cloak for plane wave
illumination \cite{Valentine09} or for building-up a curved reflector that mimics a flat mirror for SPPs \cite{Renger}. Similar ideas could be applied to the design of different devices, including the beam shifter analyzed in this work. Therefore, we are confident that the recipes that we have proposed could be practically implemented by properly structuring a dielectric layer placed on top of a metallic film at a length scale much smaller than the operating wavelength.

In conclusion, we have demonstrated how the concept of
Transformation Optics when applied to SPP propagation could offer
the necessary tools to design new strategies for molding the flow
of SPPs in structured metal surfaces. A simplification of the
general TO recipes in which only the dielectric side is
manipulated yields to almost perfect functionalities for
SPP-devices. We hope that our results will encourage other
experimental and theoretical groups to explore the exciting
opportunities that the idea of Transformation Optics brings into
the field of Plasmonics. 

Note: As we completed this manuscript, we accidentally found that another group \cite{Liu10} also completed a paper independently at the same time on a similar topic.

\acknowledgement This work has been sponsored by the Spanish
Ministry of Science under projects MAT2009-06609-C02 and
CSD2007-046-NanoLight. P.A.H. acknowledges financial support from the Spanish Ministry of Education through grant AP2008-00021. 


\newpage

\newpage

\newpage

\end{document}